\documentclass{aa}  

\usepackage{graphicx}
\usepackage{txfonts}
\usepackage{natbib}
\bibpunct{(}{)}{;}{a}{}{,}
\begin{document} 

   \title{First spectroscopic observations of the sub-stellar companion of the young debris disk star PZ Telescopii
\thanks{
Based on observations made with ESO Telescopes at the La Silla Paranal Observatory under programme ID 087.C-0109(A)
}
}

   \titlerunning{Spectroscopic observations of the sub-stellar companion of the young debris disk star PZ Telescopii}

   \author{T. O. B. Schmidt\inst{1}
           \and
           M. Mugrauer\inst{1}
	   \and
           R. Neuh\"auser\inst{1}
           \and
	   N. Vogt\inst{2}
           \and          
           S. Witte\inst{3}
           \and
           P. H. Hauschildt\inst{3}
           \and
           Ch. Helling\inst{4}
           \and
           A. Seifahrt\inst{5}
           }

   \offprints{Tobias Schmidt, e-mail:tobi@astro.uni-jena.de}

   \institute{Astrophysikalisches Institut und Universit\"ats-Sternwarte, Universit\"at Jena, Schillerg\"a\ss chen 2-3, 07745 Jena, Germany\\
              \email{tobi@astro.uni-jena.de}
         \and
              Departamento de F\'isica y Astronom\'ia, Universidad de Valpara\'iso, Avenida Gran Breta\~na 1111, Valpara\'iso, Chile
         \and
             Hamburger Sternwarte, Gojenbergsweg 112, 21029 Hamburg, Germany
         \and
             SUPA, School of Physics and Astronomy, University of St. Andrews, North Haugh, St. Andrews KY16 9SS, UK
         \and
             Department of Astronomy and Astrophysics, University of Chicago, 5640 S. Ellis Ave., IL 60637, USA
            }

   \date{Received 2013; accepted 2014}

 
  \abstract
   {In 2010 a sub-stellar companion to the solar analog pre-main sequence star PZ Tel and member of the about 12 Myr old $\beta$~Pic moving group was found by
    high-contrast direct imaging independently by two teams.}
   {In order to determine the basic parameters of this companion more precisely and independent of evolutionary models, hence age independent, we obtained follow-up spectroscopic
    observations of primary and companion.}
   {We use the Spectrograph for INtegral Field Observations in the Near Infrared (SINFONI) at the Very Large Telescope Unit 4/YEPUN of ESO's Paranal
    Observatory in H+K band and process the data using the spectral deconvolution technique. The resulting spectrum of the companion is then compared to
    a grid of {\sc Drift-Phoenix} synthetic model spectra, a combination of a general-purpose model atmosphere code with a non-equilibrium, stationary cloud and
    dust model, using a $\chi^2$ minimization analysis.}
   {We find a best fitting spectral type of G6.5 for PZ Tel A. The extracted spectrum of the sub-stellar companion,
    at a spatial position compatible with earlier orbit estimates,
    yields a temperature T$_{\rm eff}$=\,2500\,$^{+138}_{-115}$\,K, a visual extinction $A_{V}$=\,0.53\,$^{+0.84}_{-0.53}$\,mag, a surface gravity of
    $\log{g}$=\,3.50$^{+0.51}_{-0.30}$\,dex, and a metallicity at the edge of the grid of [M/H]=\,0.30$_{-0.30}$\,dex.}
   {We derive a luminosity of $\log(L_{bol}/L_{\odot})$=\,-2.66$^{+0.06}_{-0.08}$, a radius of R=\,2.42$^{+0.28}_{-0.34}$\,R$_{\mathrm{Jup}}$ and a mass of
    M=\,7.5$^{+16.9}_{-4.3}$\,M$_{\mathrm{Jup}}$ for the PZ Tel companion, being consistent with most earlier estimates using photometry alone.
    Combining our results with evolutionary models, we find a best fitting mass of about
    21 Jupiter masses at an age corresponding to the recently determined lithium depletion age of 7$^{+4}_{-2}$ Myr. Hence, the PZ Tel companion is most likely
    a wide brown dwarf companion in the 12$^{+8}_{-4}$ Myr old $\beta$~Pic moving group.}

   \keywords{Stars: brown dwarfs, pre-main sequence, atmospheres – planetary systems: formation – Stars: individual: PZ Tel}

   \maketitle
%

\section{Introduction}

In 2010 two groups performing high-contrast direct imaging surveys around young nearby stars \citep{2010A&A...523L...1M,2010ApJ...720L..82B}
independently reported the discovery of a sub-stellar companion around the young solar analog pre-main-sequence star \object{PZ Tel} (HD 174429, HIP 92680),
a likely member of the 12$^{+8}_{-4}$ Myr old $\beta$~Pic moving group \citep{2001ApJ...562L..87Z,2006A&A...460..695T} having a spectral type of
G5 -- K8 \citep{1936pmsz.book.....S,2010A&A...520A..15M} at an age of 12.8\,$\pm$\,2.2 Myr \citep{2011MNRAS.410..190T}.

While not detectable in 2003 because of its proximity to the host star \citep{2005ApJ...625.1004M}, the sub-stellar companion
was first observed on its orbit in 2007 at a
projected separation of 0.255 arcsec (13 au at 51.5\,$\pm$\,2.6 pc \citep{2007A&A...474..653V}), identified by \citet{2010A&A...523L...1M} in archival data of \citet{2010A&A...509A..52C}
after detection at 0.337 arcsec in 2009 by \citet{2010A&A...523L...1M}. \citet{2008ApJ...681.1484R} detected excess emission in the spectral energy distribution
of PZ Tel at 70 $\mu$m with MIPS/Spitzer, indicating the star is surrounded by a low-mass, cold ($\sim$41\,K) debris disk. On the basis of further astrometric
data \citet{2012MNRAS.424.1714M} could show that the preliminary determined orbital parameters of the PZ Tel companion are compatible with this disk being circumbinary
(around both components), but not with being circumstellar, and that a semimajor axis of the nearly edge-on orbit not exceeding 25 au is likely.

\begin{table*}
\caption{VLT/SINFONI observation log}
\label{table:1}
\centering
\begin{tabular}{llccccccccc}
\hline\hline
Object             & JD - 2450000      & Date of         & DIT        & NDIT   & Number    & Airmass & DIMM$^a$   & $\tau_{0}^b$ \\
                   & $[\mathrm{days}]$ & observation     & [s]        &        & of images &         & Seeing     & [ms]         \\
\hline
PZ Tel             & 5795.62984        & 22 Aug 2011     & 5          & 9      & 16        & 1.17    & \ \, 1.35   & 1.2          \\
\object{HIP 94378} & 5795.65222        & 22 Aug 2011     & 2          & 5      & 1         & 1.11    & $\sim$1.49 & 1.1          \\
\hline
\end{tabular}
\begin{flushleft}
\textbf{Remarks}: All observations were done in H+K band and 0.025 mas/spaxel scale (FoV: 0.8 arcsec x 0.8 arcsec).
(a) Differential image motion monitor (DIMM) Seeing average of all images (b) coherence time of atmospheric fluctuations. 
\end{flushleft}                                                                                                     
\end{table*}

\begin{figure*}
\centering
\resizebox{0.8\textwidth}{!}{\includegraphics{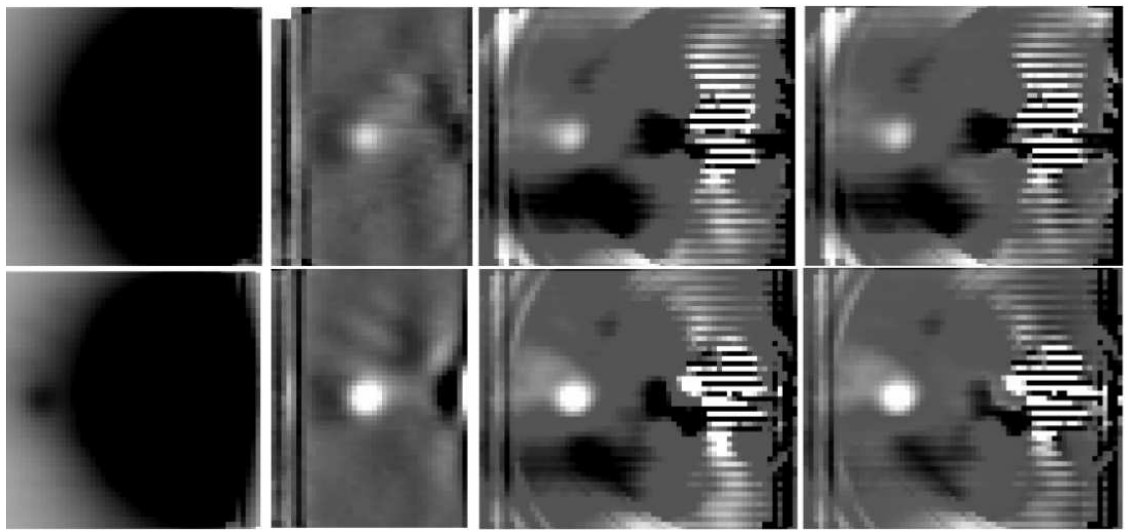}}
\caption{Averaged data cubes obtained of the PZ Tel companion in H-band (\textit{upper row}) and K-band (\textit{lower row}) at position angle -30$^\circ$.
\textit{From left to right in each row}: Cube after data reduction, eastern part of cube after 2 dimensional polynomial fit of the primary PSF and removal of the primary halo,
cube after subtration of the radial symmetric part of the primary PSF, cube after subtration of the radial symmetric part of the primary PSF and additional spectral
deconvolution. For clarity flux is inverted in the latter
3 images of each row. See text for details.}
\label{FigPSFRemoval}
\end{figure*}

Both \citet{2010ApJ...720L..82B} using NICI at Gemini-South and \citet{2010A&A...523L...1M} using NACO at ESO VLT concluded from photometric data in their
discovery papers that the imaged object is likely a brown dwarf companion to the $\beta$~Pic moving group member PZ Tel.
\citet{2010ApJ...720L..82B} find best fitting parameters of 2702\,$\pm$\,84 K and log g of 4.20\,$\pm$\,0.11 dex, hence 36\,$\pm$\,6 M$_{\mathrm{Jup}}$, while
\citet{2010A&A...523L...1M} find 2500--2700 K, hence 28$^{+12}_{-4}$ M$_{\mathrm{Jup}}$, both using evolutionary models \citep{2000ApJ...542..464C}.
\citet{NeuhSchmidt2012} noted that given this lower mass limit, the PZ Tel companion could even be a planetary mass object, considering the planet definition by
\citet{2011A&A...532A..79S}. Recently, \citet{2012MNRAS.420.3587J} derived an age of the system of 24\,$\pm$\,3 Myr, based on
evolutionary models and chromospheric activity of the primary,
contrasting with their own age determination of 7$^{+4}_{-2}$ Myr obtained from lithium depletion.
This higher age estimate yields a higher effective temperature 2987\,$\pm$\,100 K, surface gravity ($\log{g}$) 4.78\,$\pm$\,0.10 dex and mass 62\,$\pm$\,9
M$_{\mathrm{Jup}}$. They moreover determined the metallicity [Fe/H] of the system to be 0.05\,$\pm$\,0.20 dex.

In order to determine the basic parameters of the PZ Tel companion more precisely and independent of evolutionary models we obtained follow-up spectroscopy.
Here we present our results from these observations, targeted to identify the nature of the sub-stellar companion.

\section{Observations and data reduction}

We used the adaptive optics integral-field spectrograph SINFONI, mounted at UT4 of the ESO VLT.
The observations of the PZ Tel companion were carried out in H+K band (resolution 1500) using the best resolving 0.0125 mas/spaxel scale of the instrument
with a FoV of 0.8\arcsec\,$\times$\,0.8\arcsec. Further details are summarized in Table~\ref{table:1}.

We used the SINFONI data reduction pipeline version 2.0.5 offered by ESO \citep{2006ASPC..351..295J} with reduction routines developed by the
SINFONI consortium \citep{2006NewAR..50..398A}. After standard reduction, including dark subtraction, flat fielding, distortion correction and wavelength
calibration, all nodding cycles were combined to a final data cube.

We used the \textit{Starfinder} package of IDL \citep{2000SPIE.4007..879D} and an updated version of an algorithm to remove the halo of PZ Tel
by 2 dimensional polynomial fit of the primary PSF, both described
in detail in \citet{2007A&A...463..309S} and \citet{2008A&A...491..311S}. The data quality of July 2011 was too bad for the data to be used in general,
however for the data of August 2011 we get the result presented as the second image in each row in Fig.~\ref{FigPSFRemoval}.
The technique was developed for wider companions and the primary being off the FoV, thus an overshooting polynomial removal is present after application.

Since the primary is in the FoV this additional information can be used for the removal of the primary halo.
We use a technique called spectral deconvolution introduced by \citet{2002ApJ...578..543S} for HST data and used for the first time for ground-based data
by \citet{2007MNRAS.378.1229T}. The idea is to remove speckles and other remnants, which move with wavelength across the cube and can be distinguished from the
non-moving companion, by rescaling the cube according to the wavelength of each plane in the cube. Then the companion is at different positions at different
wavelength/spectral channels in the new cube, while wavelength dependent speckles and remnants remain at the same positions and can be identified using a median and removed
from the original cube. Please see \citet{2007MNRAS.378.1229T} for details of the technique. It, however, only works if the wavelength coverage and the projected
separation of the two components have an appropriate ratio to each other, ensuring that the companion moves enough after rescaling to remove its PSF by
median combination of a small fraction of wavelength channels at either end of the data cube. Such a clean collapsed image devoid of flux from the companion can
be constructed in our case as the $\epsilon$ value defined in \citet{2007MNRAS.378.1229T} easily exceeds unity, being 1.83 for the projected separation of
PZ Tel \& its sub-stellar companion (see next section).

In a first step we have to remove the strongest part of the influence of the halo of the primary, as the halo full width strongly changes with wavelength because of the
difference in adaptive optics (AO) performance at different wavelength. We achieve that by determination and removal of the radial symmetric part of the primary PSF,
as can be seen in the third image of each row in Fig.~\ref{FigPSFRemoval}. The center determination and reshifting of the PSF was improved even further to the results
of \textit{Starfinder} \citep{2000SPIE.4007..879D} by fitting a polynomial of 2nd degree to the x shift values of the primary PSF from spectral channel to spectral
channel in the cube to suppress any rapid changes on short
wavelength differences. This can be neglected for y shift values as the cube was aligned in y according to the position angle of $\sim$~60$^\circ$ of
the sub-stellar companion with respect to the primary for the observations.

We then perform the spectral deconvolution technique using a self-written version in IDL.
To fully optimize the technique we also remove the PSF of the secondary within an iterative spectral deconvolution implementation in 10 iteration steps, as similarly
done in \citet{2007MNRAS.378.1229T} and arrive at the final images of each row in Fig.~\ref{FigPSFRemoval}. As can be seen, not all remnants could be removed. We
note that this is expected as the reproducibility of speckles and similar remnants and hence the efficiency of the technique decreases with increasing Seeing and
decreasing coherence time, both not ideal in our observations (Table \ref{table:1}). 

As telluric standard HIP 94378, a B5V star was used for H+K band. In order to correct for features of this standard star, the Bracket-$\gamma$ line in the K band
as well as the $\zeta$, $\theta$, $\eta$ and $\iota$ lines of the Bracket series in the H band were removed before usage of the standard. A correction temperature of
15400 K was assumed for the star, as given for a B5V in \citet{1995ApJS..101..117K}.
We use the optimal extraction technique by \citet{1986PASP...98..609H} to improve the quality of the standard, as it was observed at a slightly worse Seeing
(Table \ref{table:1}).  

We cannot use PSF fitting of the sub-stellar companion as the S/N is quite low in parts of the H band, hence aperture spectroscopy is our choice.
However, we realized that the spectral shape changes with different aperture sizes, a natural outcome of different Strehl ratios achieved at the different
wavelength used. Hence, we apply a correction fuction to the extracted spectrum: We extract the spectrum of the primary at the same aperture size as the companion
and at the maximum possible aperture size and multiply the companion spectrum by the quotient of these spectra to correct for the differences of AO performance
achieved at different wavelengths. The optimal extraction is not used in order not to emphasize artificial deviations.

Without the correction function we find temperatures for the companion about 700 K lower than our final result. Nevertheless different aperature sizes still give
different results, which can be attributed to different AO performances achieved for primary and companion and/or remnants in the final cubes not removed by
reduction and spectral deconvolution. Because of the proximity of the companion the AO performance at primary and companion position should be very comparable.
Since we see in addition strong and spatially close positive remnants in the K band, artificially increasing the flux for increasing aperture sizes, we conclude that
these remnants are the dominant source of changes in the spectral shape in accordance with aperture size. Hence, the accuracy of the spectrum should increase with
decreasing aperture size, while its precision gets worse. 

At the beginning of the H band at 1.45 $\mu$m, the sparrow resolution limit \citep{1916ApJ....44...76S} is at 34.6 mas or 2.8 spaxels, hence we choose an aperture
size of 2.5 spaxels. This choice guarantees the best accuracy (without further inclusion of remnants), while a lower value would be below the resolution and comes
into conflict with the precision of spatial PSF alignment at different wavelengths.

\section{Results}

\subsection{Astrometry}

An average projected separation of 32 spaxels was found between the components of the PZ Tel system.
As no astrometric calibration of SINFONI was observed, we cannot give accurate numbers.
However, 32 spaxels correspond to 0.4 arcsec using the nominal spaxel scale of 0.0125 \arcsec/spaxel,
being well in agreement with the orbit determination of \citet{2012MNRAS.424.1714M}.

\subsection{Spectroscopy of PZ Tel}

\begin{figure}
\centering
\resizebox{0.5\textwidth}{!}{\includegraphics{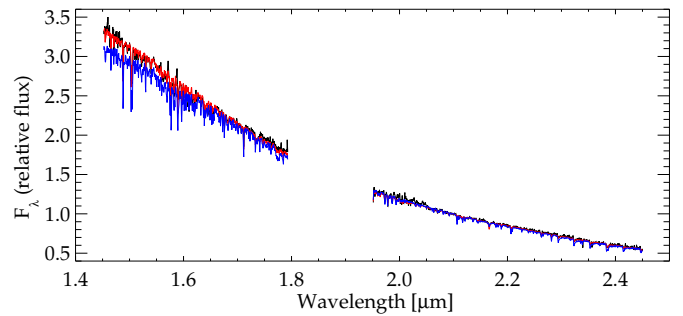}}
\caption{Standard calibrated spectrum of PZ Tel (black) in comparison to 
an IRTF Spectral Library G6.5V comparison spectrum (red) and a K0V comparison spectrum (blue).
See text for details.}
\label{FigPrimSpectrum}
\end{figure}

As PZ Tel is within the FoV, dominating the obtained flux, we can extract a spectrum of high S/N.
Fig.~\ref{FigPrimSpectrum} shows the standard calibrated spectrum of PZ Tel (black) in comparison to 
IRTF (NASA Infrared Telescope Facility) Spectral Library products \citep{2009ApJS..185..289R}. A best fit
is achieved for the G6.5V comparison spectrum (red) from the spectral library, while a K0V (blue), as found
by \citet{1993yCat.3051....0H} and commonly given as spectral type \citep{1997ESASP1200.....P},
exhibits a too red spectral continuum. G6.5V corresponds to about 5665 K \citep{1995ApJS..101..117K},
which agrees well with the temperature estimates 5623 K in \citet{1999A&A...352..555A} and 5308--6065 K in
\citet{2011MNRAS.411..435B}, as well as earlier spectral type determinations as G5 \citep{1936pmsz.book.....S}.
Discrepancies could be attributed to variability, common at the youth of PZ Tel.

\subsection{Spectroscopy of the PZ Tel companion}

\begin{figure*}
\centering
\resizebox{\textwidth}{!}{\includegraphics{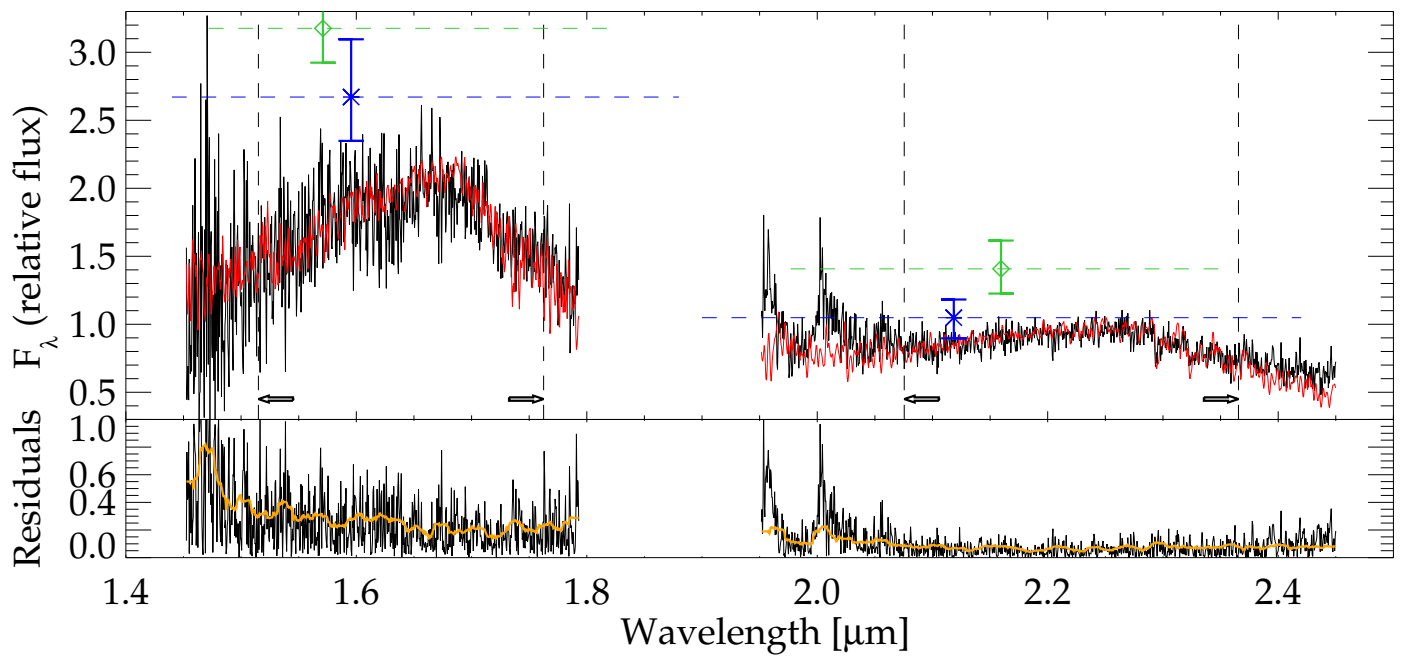}}
\caption{\textit{Top:} Our SINFONI spectrum of the PZ Tel companion (black) in spectral resolution 1500 in comparison to the best fitting
{\sc Drift-Phoenix} synthetic spectrum (red; same spectral resolution) of T$_{\rm eff}$=\,2500\,K, $\log{g}$=\,3.5\,dex, [M/H]=\,0.3\,dex and a
visual extinction of $A_{V}$=\,0.53\,mag.
The spectral range used for $\chi^2$ minimization is indicated by arrows as well as vertical dashes lines.
Magnitude differences with regard to the primary as given in \citet{2010A&A...523L...1M} (blue) and \citet{2010ApJ...720L..82B} (green) were
transformed to relative fluxes using the NACO and NICI/NIRI filter transmission curves, respectively.
The spectral range of the filters is given as dashed lines, in contrast to negligible spectral error bars.
\textit{Bottom:} Absolute value of the residuals of the PZ Tel companion spectrum and the best model fit (black) in comparison to the determined noise floor (orange).
See text for details.}
\label{FigSpectrum}
\end{figure*}

As described in the previous section extensive reductions and corrections had to be applied to improve the quality of the extracted spectrum, including
standard reduction, rotation subtraction, spectral deconvolution, application of an AO performance correction function as well as best choice of aperture size. 

As dust begins to condensate at temperatures T$_{\mathrm{eff}}$\,$\approx$\,2700 K \citep{1996A&A...308L..29T,2005MNRAS.361.1323C}, we compare the spectrum of
the sub-stellar companion to synthetic model spectra taking the influence of dust within a kinetic model description of the formation and
evolution of dust in brown dwarfs
into account \citep{2008ApJ...675L.105H}. The employed {\sc Drift-Phoenix} models combine a non-equilibrium, stationary cloud model from
\citet[][{\sc DRIFT}: Nucleation, seed formation, growth, evaporation, gravitational settling, convective overshooting\,/\,up-mixing,
element conservation]{2008A&A...485..547H} with a general-purpose model atmosphere code \citep[][{\sc PHOENIX}: Radiative transfer, hydrostatic
equilibrium, mixing length theory, chemical equilibrium]{1999JCoAM.109...41H}.

In contrast to \citet{2008A&A...491..311S} we use {\sc Drift-Phoenix} v1.2, a complete grid of models with new equation of state in the range of 
T$_{\rm eff}$=\,1000\,$\ldots$\,3000\,K, $\log{g}$=\,3.0\,$\ldots$\,5.5 dex, and [M/H]=\,-0.6\,$\ldots$\,0.3 dex in steps of 100\,K, 0.5\,dex, and 0.3\,dex,
respectively. Moreover, we still need to account for reddening of our spectra by extinction. This correction is important as the effective temperature is
highly correlated with the extinction, because both values change the slope of the spectrum in H+K band.

In order to achieve more precise best fitting values we linearly interpolated the present {\sc Drift-Phoenix} v1.2 grid to step sizes of
10\,K, 0.1\,dex, and 0.1\,dex in temperature, logarithm of surface gravity and logarithm of metallicity, respectively.

We used a $\chi^2$ minimization algorithm to find the best-fitting combination of (a) the effective temperature, (b) the surface gravity, (c) the metallicity,
and (d) the extinction correction of the measured spectrum, a version similar to the one used in \citet{2008A&A...491..311S}, but strongly improved in
usability and accuracy, by e.g.~a more precise determination of the noise floor of the data, see Fig.~\ref{FigSpectrum}.

In the upper panel of Fig.~\ref{FigSpectrum} we give the best fit of our spectrum (black) with a reduced $\chi^2_{red}$ of 1.16.
The (best) fitting {\sc Drift-Phoenix} model (red) of the companion of PZ Tel has a temperature of T$_{\rm eff}$=\,2500\,$^{+138}_{-115}$\,K, a visual extinction
$A_{V}$=\,0.53\,$^{+0.84}_{-0.53}$\,mag, a surface gravity of $\log{g}$=\,3.50$^{+0.51}_{-0.30}$\,dex, and a metallicity of [M/H]=\,0.30$_{-0.30}$\,dex. An upper
limit for the metallicity cannot be given, as the best fitting value is at the upper edge of the used model grid.
To improve on the accuracy we linearly fitted 1, 2, and 3 $\sigma$ values to determine the final 1 $\sigma$ error.
In addition we give the absolute value of the residuals of the fit (black) in the lower panel of Fig.~\ref{FigSpectrum} in comparison to the determined noise floor of the spectrum (orange).

The spectral range used for the $\chi^2$ minimization is indicated by arrows as well as vertical dashes lines in Fig.~\ref{FigSpectrum}.
In this area the average S/N is 7.5 for the H- and 12.6 per spaxel for the K-band.
At a resolution of 1500 in H+K we derive an average spectral scale of about 2.5 spaxels per resolution element. Using this value we determine 
$\sim$423 of 1077 data points to be independent spectral measurements. On this basis we can calculate the $\chi^2$ values corresponding to 
1, 2 and 3 $\sigma$ significance of 423 - 4 (4 parameters to fit) degrees of freedom to be 432, 469 and 504, respectively, and after correction for the full
number of measured spaxels to be 1100, 1194 and 1283 for 1, 2 and 3 $\sigma$, respectively.

\begin{figure}
\centering
\resizebox{0.5\textwidth}{!}{\includegraphics{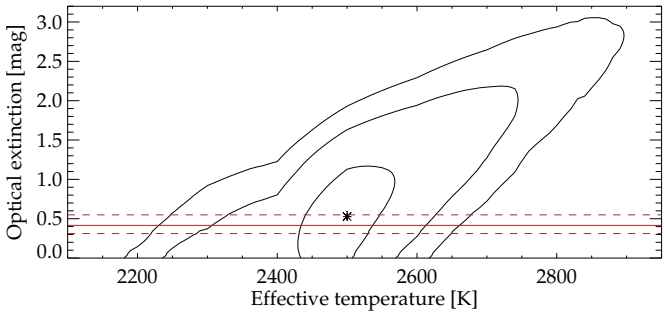}}
\caption{Result of the $\chi^2$ minimization analysis for the PZ Tel companion. Plotted are the best value (\textit{asterisk}) and the 1, 2, and 3 sigma error
contours for effective temperature T$_{\rm eff}$ and optical extinction $A_{V}$, determined from comparison of our SINFONI spectrum and the {\sc Drift-Phoenix} model grid.
Further the visual extinction of PZ Tel A is shown (horizontal brown solid line), including its 1 sigma errors (dashed lines) as found
by \citet{2011MNRAS.411..435B}.}
\label{FigContourExtinction}
\end{figure}

\begin{figure}
\centering
\resizebox{0.5\textwidth}{!}{\includegraphics{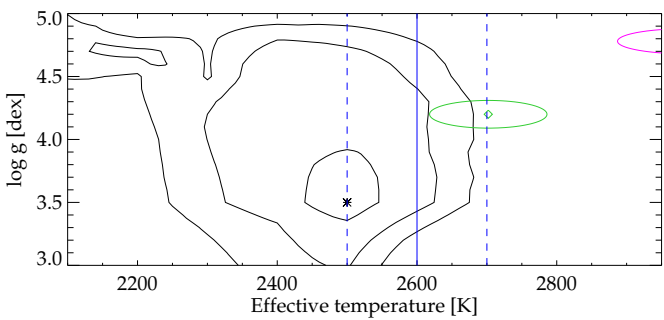}}
\caption{Result of the $\chi^2$ minimization analysis for the PZ Tel companion. Plotted are the best value (\textit{asterisk}) and the 1, 2, and 3 sigma error
contours for effective temperature T$_{\rm eff}$ and logarithm of the surface gravity $\log{g}$, determined from comparison of our SINFONI spectrum and the
{\sc Drift-Phoenix} model grid. Further the temperature is shown as found by \citet{2010A&A...523L...1M} (vertical blue solid line), including
its 1 sigma errors (dashed lines), surface gravity and 
temperature as found by \citep{2010ApJ...720L..82B} (green ellipse) and by \citet{2012MNRAS.420.3587J} (magenta part of ellipse).}
\label{FigContourSurfaceGravity}
\end{figure}

\begin{figure}
\centering
\resizebox{0.5\textwidth}{!}{\includegraphics{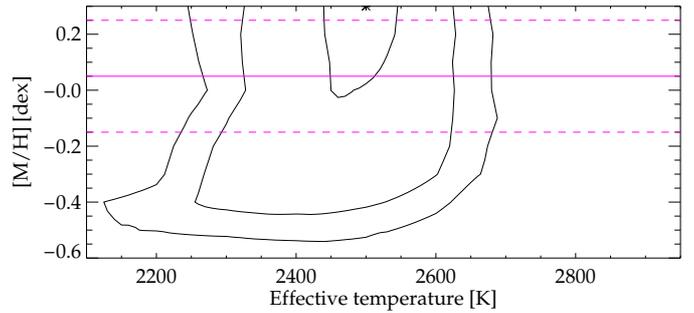}}
\caption{Result of the $\chi^2$ minimization analysis for the PZ Tel companion. Plotted are the best value (\textit{asterisk}) and the 1, 2, and 3 sigma error
contours for effective temperature T$_{\rm eff}$ and metallicity [M/H], determined from comparison of our SINFONI spectrum and the {\sc Drift-Phoenix} model grid.
Further the metallicity of PZ Tel A is shown as found by \citet{2012MNRAS.420.3587J} (horizontal magenta solid line), including its
1 sigma errors (dashed lines).}
\label{FigContourMetallicity}
\end{figure}

These significances are given as contours in Fig.~\ref{FigContourExtinction} for temperature vs.~visual extinction, in Fig.~\ref{FigContourSurfaceGravity}
for temperature vs.~surface gravity and in Fig.~\ref{FigContourMetallicity} for temperature vs.~metallicity.
In addition comparison values from the literature are given as straight lines with 1 $\sigma$ error bars as dashed lines, color coded for different authors or
as 1 $\sigma$ error ellipses if both values plotted were given.
In brown extinction of the primary PZ Tel by \citet{2011MNRAS.411..435B}, in magenta temperature, surface gravity and metallicity (of the primary) as found by
\citet{2012MNRAS.420.3587J}, in blue temperature as found by \citet{2010A&A...523L...1M} and in green temperature and surface gravity as found  by
\citet{2010ApJ...720L..82B}. All comparison values were determined using photometry and evolutionary models. The extinction value of PZ Tel A had to be
transformed to visual extinction using the formula given in \citep{2011MNRAS.411..435B} as well as to 1 $\sigma$ significance.

\subsection{Photometry}

In the upper panel of Fig.~\ref{FigSpectrum} photometric data points from \citet{2010ApJ...720L..82B} (green) and \citet{2010A&A...523L...1M} (blue) are
given for comparison. The magnitude differences given were transformed into flux differences in regard to PZ Tel A by convolving the spectra of PZ Tel and
its sub-stellar companion with the filter transmission curves in H- and Ks-band of NIRI, same filters as used for the Gemini NICI planet-finding campaign
\citep{2010SPIE.7736E..53L}, and the transmission curves of NACO \citep{2003SPIE.4841..944L,2003SPIE.4839..140R}, respectively. As output of the convolution
we determine the corresponding fluxes of the previously conducted photometric observations as well as the effective spectral center of the measurements at which
the points are drawn, which differ slightly from the midpoint of the respective filter transmission curves, whose FHWM are shown as horizontal dashed lines
in Fig. \ref{FigSpectrum}. Note that the full spectral range
observed is 1.45 -- 2.45 $\mu$m, used for the convolution, but not shown due to heavy water vapour influence in the case of the challenging observing conditions
(Table~\ref{table:1}).
The H-Ks color of the companion is constant across all measurements within 0.8 $\sigma$ errors. 

In Table~\ref{table:2} we present the measured flux differences between PZ Tel and its sub-stellar companion.
These differences were measured as described in the previous passage, with the
single difference that now flux differences were transformed into magnitude differences by the same procedure, i.e.~convolving the spectra of
PZ Tel and its sub-stellar companion
with the filter transmission curves in H- and Ks-band of NIRI, same filters as used for the Gemini NICI planet-finding campaign
\citep{2010SPIE.7736E..53L}, and the transmission curves of NACO \citep{2003SPIE.4841..944L,2003SPIE.4839..140R}, respectively.
In order to compute an H-Ks color of the PZ Tel companion, we assumed H-Ks of PZ Tel to be as measured by
2MASS \citep{2003tmc..book.....C,2006AJ....131.1163S}. 

In addition we give the maximum possible systematic errors in Table~\ref{table:2}, which are present if the strongest positive/negative remnant features in Fig.~\ref{FigPSFRemoval}
are superimposed onto the PZ Tel companion spectrum. This amount of remnants is, however, unlikely. On the one hand there is no evident deformation of the sub-stellar companion's PSF, except
a brightening towards northeast (see Fig.~\ref{FigPSFRemoval}), which, however, would decrease the magnitude difference between PZ Tel and its sub-stellar companion in the current period (see
Table \ref{table:2}). On the other hand our H-Ks color after i.a.~spectral deconvolution and AO performance correction, but without the consideration of systematic errors,
is only deviant by 0.7 \& 0.2 $\sigma$ with regard to the results
by \citet{2010A&A...523L...1M} \& \citet{2010ApJ...720L..82B}, respectively (Table \ref{table:2}).

\begin{table}
\caption{PZ Tel companion photometry and magnitude differences}
\label{table:2}
\centering
\begin{tabular}{lccc}
\hline\hline
Source         & $\Delta$H & $\Delta$Ks & H-Ks  \\ 
               & [mag]     & [mag]      & [mag] \\
\hline
Biller (a,b)   & 5.38 $\pm$ 0.09 & 5.04 $\pm$ 0.15 & 0.45 $\pm$ 0.18 \\
here (b)       & 5.90 $\pm$ 0.16 & 5.52 $\pm$ 0.09 & 0.51 $\pm$ 0.19 \\
Mugrauer (c,d) & 5.51 $\pm$ 0.09 & 5.34 $\pm$ 0.06 & 0.29 $\pm$ 0.12 \\
here (d)       & 5.91 $\pm$ 0.20 & 5.55 $\pm$ 0.11 & 0.48 $\pm$ 0.23 \\
\hline
maximum sys-   &  + $\leq$ 0.7 &  + $\leq$ 0.8 &  \\
tematic error  & -- $\leq$ 0.8 & -- $\leq$ 0.5 &  \\
\hline
\end{tabular}
\begin{flushleft}
\textbf{Remarks}: (a) \citet{2010ApJ...720L..82B}, (b) NICI/NIRI filters,
(c) \citet{2010A&A...523L...1M}, (d) NACO filters
\end{flushleft}                                                                                                     
\end{table}

\section{Mass determination and conclusions}

Using the photometry of PZ Tel A from the Two Micron All Sky Survey (2MASS) catalog \citep{2003tmc..book.....C,2006AJ....131.1163S}
of $K$\,=\,6.366\,$\pm$\,0.024\,mag we can estimate all further parameters of the sub-stellar companion using all results from the spectroscopic
analysis described in the previous section. As no absolute photometric calibration is possible with the spectroscopic standard,
we preliminary estimate the parameters of the PZ Tel companion assuming negligible photometric variability of both sources, most likely not 
correct according to variability indications presented in the previous sections.

We derive a luminosity of $\log(L_{bol}/L_{\odot})$=\,-2.66$^{+0.06}_{-0.08}$ for the PZ Tel companion from the extinction corrected
apparent brightness Ks$_{0}$= 11.86$^{+0.07}_{-0.10}$\,mag \cite[from the 2MASS brightness, the magnitude difference (Table~\ref{table:2}),
$A_{V}$=\,0.53\,$^{+0.84}_{-0.53}$\,mag and extinction law by][]{1985ApJ...288..618R} using a bolometric correction of
B.C.$_{K}$\,=\,3.1\,$\pm$\,0.1\,mag from \citet[][for spectral type M6--L0]{2004AJ....127.3516G} at a distance of
51.49$^{+2.74}_{-2.47}$\,pc \citep{2007A&A...474..653V}.
From the luminosity and temperature T$_{\rm eff}$=\,2500\,$^{+138}_{-115}$\,K, we calculate the radius to be
R=\,0.25$^{+0.03}_{-0.04}$\,R$_{\odot}$ or 2.42$^{+0.28}_{-0.34}$\,R$_{\mathrm{Jup}}$.
From radius and surface gravity $\log{g}$=\,3.50$^{+0.51}_{-0.30}$\,dex, we find a mass of the PZ Tel companion of
M=\,0.0071$^{+0.0161}_{-0.0041}$\,M$_{\odot}$ or 7.5$^{+16.9}_{-4.3}$\,M$_{\mathrm{Jup}}$.

For the bolometric correction we used the spectral type corresponding to the temperature range found here. Using the H$_2$O index defined
in \citet{2007ApJ...657..511A} we find a possible spectral range of L1.5 -- L4 for the sub-stellar companion, using the modified version in \citet{2014A&A...562A.127B}
we find a possible spectral range of M7 -- L9, roughly consistent with our findings of M6 -- L0. \citet{2014A&A...562A.127B} modified the index to avoid noisy
regions, definitely necessary in the present case, as the blue part of the H-band has the lowest signal to noise.
Our derived surface gravity uncertainties are about equal to the values of 0.5 dex, as computed by \citet{2014A&A...562A.127B} for similar objects.

The derived values, diagramed in Figs.~\ref{FigContourExtinction}--\ref{FigContourMetallicity}, agree within 1 $\sigma$
with the extinction $A_{V}$ and metallicity [M/H] for PZ Tel A by \citet{2011MNRAS.411..435B} and \citet{2012MNRAS.420.3587J}, respectively
as well as with the temperature by \citet{2010A&A...523L...1M}. Temperature and surface gravity deviate in regard to literature estimates
by 1.2 \& 1.3 $\sigma$ \citep{2010ApJ...720L..82B} and 2.9 \& 2.5 $\sigma$ \citep{2012MNRAS.420.3587J}, respectively. Finally our mass
result deviates by 1.2 $\sigma$ \cite[28$^{+12}_{-4}$\,M$_{\mathrm{Jup}}$,][]{2010A&A...523L...1M}, by 1.6 $\sigma$
\cite[36\,$\pm$\,6\,M$_{\mathrm{Jup}}$,][]{2010ApJ...720L..82B} and by 3.2 $\sigma$ \cite[62\,$\pm$\,9\,M$_{\mathrm{Jup}}$,][]{2012MNRAS.420.3587J}.

Although our mass estimate is independent of evolutionary models, we can use them as comparison to put our results into context and to check which
age is indicated by the models for our spectroscopic results. According to \citet{2000ApJ...542..464C} DUSTY models best fits are achieved
between 5 -- 10 Myr isochrones, fitting our spectral results for temperature, surface gravity and luminosity within 1 $\sigma$ errors.
This age range is consistent with the age of the $\beta$~Pic moving group (12$^{+8}_{-4}$ Myr), deviant from the age determined by
\citet{2012MNRAS.420.3587J} (24\,$\pm$\,3 Myr), while very well consistent with the lithium depletion age of PZ Tel A of 7$^{+4}_{-2}$ Myr
by the same authors. Very recently \citet{2014MNRAS.438L..11B} combined data of eight low-mass candidates with literature data of $\beta$~Pic moving group members and find
a lithium depletion age of 21\,$\pm$\,4 Myr, being also inconsistent with the age found here.

Finally we arrive at a possible mass range of 3.2 -- 24.4 M$_{\mathrm{Jup}}$.
According to these estimates the PZ Tel companion is most likely a brown dwarf of about
21 Jupiter masses, as the evolutionary models reject a $\log{g}$\,=\,3.5\,dex to be only valid for objects younger than 1 Myr and predict a
surface gravity $\log{g}$\,$\sim$\,3.95\,dex for the given lithium depletion age range,
being within the 1 $\sigma$ uncertainty of our results.
However, spectra at improved observing conditions and with longer spectral range, especially including alkali metal
lines for a more precise surface gravity determination should be able to narrow down the parameters of the PZ Tel companion in the future.
Fortunately such spectra are increasingly easy to acquire because of the strong orbital separation increase, probably persistent for the upcoming
years to decades \citep{2012MNRAS.424.1714M}.

\begin{acknowledgements}
We would like to thank the ESO Paranal Team for carrying out the observations as well as the ESO User Support department
and all the other very helpful ESO services as well as the anonymous referee for helpful comments.\\

We would further like to thank Marlies Meyer for helpful discussions of the atmospheric physics and models.\\

TOBS \& RN would like to acknowledge support from the German National Science 
Foundation (Deutsche Forschungsgemeinschaft, DFG) in grant NE 515/30-1.
NV acknowledges support by Comit\'e Mixto ESO-Gobierno de Chile, as well as by the Gemini-CONICYT fund 32090027.
ChH highlights financial support of the European Community under the FP7 by an ERC starting grant.\\

This publication makes use of data products from the Two Micron All Sky Survey, which is a joint project of the University of
Massachusetts and the Infrared Processing and Analysis Center/California Institute of Technology, funded by the National 
Aeronautics and Space Administration and the National Science Foundation.
This research has made use of the VizieR catalog access tool and the Simbad database, both operated at the Observatoire 
Strasbourg.
This reasearch makes use of the Hipparcos Catalogue, the primary result of the Hipparcos space astrometry mission, undertaken
by the European Space Agency.
This research has made use of NASA's Astrophysics Data System Bibliographic Services.
\end{acknowledgements}

\bibliographystyle{aa}

\listofobjects

\end{document}